\documentclass[journal,draftcls,onecolumn,12pt,twoside]{IEEEtranTCOM}

\normalsize

\usepackage{amssymb}
\usepackage{mathrsfs}

\usepackage{cite}
\ifCLASSINFOpdf
  \usepackage[pdftex]{graphicx}
  \graphicspath{{../pdf/}{../jpeg/}}
  \DeclareGraphicsExtensions{.pdf,.jpeg,.png}
\else
  \usepackage[dvips]{graphicx}
  \graphicspath{{../eps/}}
  \DeclareGraphicsExtensions{.eps}
\fi
\usepackage[cmex10]{amsmath}
\usepackage{algorithm}
\usepackage{algpseudocode}
\usepackage{array}
\ifCLASSOPTIONcompsoc
  \usepackage[caption=false,font=normalsize,labelfont=sf,textfont=sf]{subfig}
\else
  \usepackage[caption=false,font=footnotesize]{subfig}
\fi

\newtheorem{myprop}{Proposition}

\begin{document}
\title{On Buffer-Aided Multiple-Access Relay Channel}

\author{Rongkuan~Liu,~\IEEEmembership{}
        Petar~Popovski,~\IEEEmembership{Fellow,~IEEE,}
        and~Gang~Wang,~\IEEEmembership{Member,~IEEE}
\thanks{R. Liu is with the Communication Research Center, Harbin Institute of Technology, Harbin 150001, China, and also with the Department of Electronic Systems, Aalborg University, Aalborg 9220, Denmark (e-mail: liurongkuan@hit.edu.cn).}
\thanks{P. Popovski is with the Department of Electronic Systems, Aalborg University, Aalborg 9220, Denmark (e-mail: petarp@es.aau.dk).}
\thanks{G. Wang is with the Communication Research Center, Harbin Institute of Technology, Harbin 150001, China (e-mail: gwang51@hit.edu.cn).}}


\maketitle

\begin{abstract}
The paper treats uplink scenario where $M$ user equipments (UEs) send to a Base Station (BS), possibly via a common Relay Station (RS) that is equipped with a buffer. This is a multiple-access relay channel (MARC) aided by a buffer. We devise a protocol in which the transmission mode is selected adaptively, using the buffer at the RS in order to maximize the average system throughput. We consider the general case in which the RS and the BS can have limits on the maximal number of transmitters that can be received over the multiple access channel. In each slot there are three type possible actions: (A1) multiple UEs transmit at rates that enable BS to decode them (A2) multiple UEs transmit, the BS can only decode the messages partially, while the RS completely; (A3) RS forwards the side information to BS about the partially decoded messages, which are going to be combined and decoded entirely at the BS, while simultaneously a number of UEs sends new messages to the BS. The results show that the adaptive selection of direct and buffer-aided relay transmissions leads to significant average throughput gains.
\end{abstract}

\begin{IEEEkeywords}
Multiple-access relay channel, buffer-aided relaying,  partial decode-and-forward.
\end{IEEEkeywords}

\section{Introduction}
The multiple-access relay channel (MARC) is a network topology where multiple user equipments (UEs) communicate with a single Base Station (BS) in the presence of a Relay Station (RS) \cite{866330}. The achievable rate region for white Gaussian MARC is investigated in \cite{1381917} and \cite{1499041} by cooperative strategies under non-phase fading and ergodic phase-fading. Particularly in \cite{1381917}, due to the constraint of half-duplex MARC, two different achievable cooperative strategies are proposed for \emph{constrained MARC}, namely (1) Decode-and-Forward (DF) and (2) Partial Decode-and-Forward (PDF). In DF strategy, the sources do not send new messages in the transmit state but simply cooperate with the relay to aid the destination in decoding the messages sent in the receive state. While in PDF strategy, in addition to cooperating with the relay in the transmit state, the sources directly transmit new messages to the destination. The results show the PDF strategy is better than the DF strategy in constrained MARC. However, the above strategies are based on fixed time scheduling, which may not take the full advantages of the channel variations. Moreover, in ergodic fading channel, it is not always necessary for the relay to decode the information. Finally, it is assumed that RS and BS can decode arbitrary number of transmitters over the respective MAC channel, as long as their rates are within the capacity region. In practice, the maximal number of transmitters may be limited due to e.g. synchronization issues.

\begin{figure}[!t]
\centering
\includegraphics[width=2.5in]{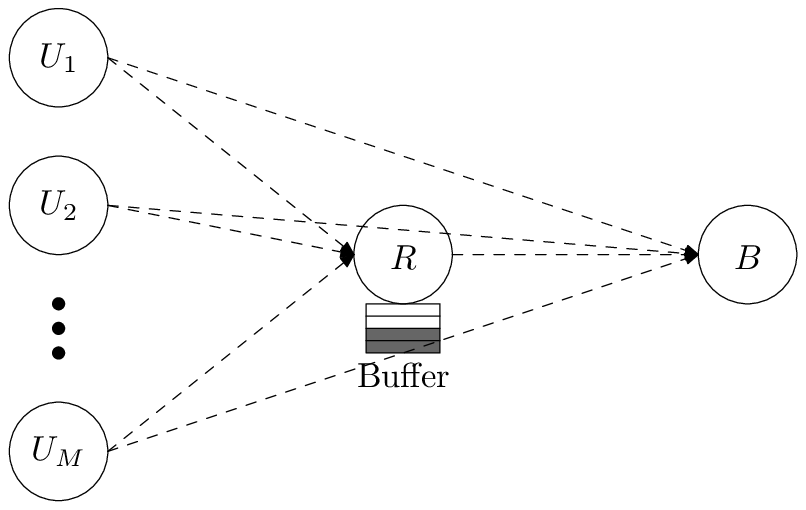}
\caption{Multiple-access relay channel with a shared buffer.}
\label{model}
\end{figure}

Here we enrich the MARC model by equipping the relay with a buffer \cite{6330084}, in order to take advantage of the channel variability through a suitable selection of the transmitters. Buffer-aided relaying for multiuser systems has been treated in \cite{6666121} where the direct links are ignored and all transmissions are orthogonalized. As MARC suggest, the direct link is an important element for achieving a better performance and only a few works have considered it in the case of buffer-aided relaying, such as \cite{6284666}, where direct and cooperative transmission are studied in a three-node network. The recent work \cite{7039198} considers a scheme that uses the direct link opportunistically along with buffer-aided relaying. There is also a line of work that studies MARC from queueing perspective \cite{R1}-\cite{R5}, where the dynamics arises from random access \cite{R1} or bursty arrivals \cite{R2} \cite{R5}, while the dynamics in our model is only driven by the channel dynamics and user selection.

In this letter we study the general MARC with a shared buffer under ergodic fading, where the RS and BS can handle $K_{R}$ and $K_{B}$ transmitters, respectively, where $K_{R} \in \{ 1,\ldots,M \}$ and  $K_{B} \in \{ 1,\ldots,M+1 \}$. It is reasonable to have $K_{B} \geq K_{R}$, while the achievable strategies in \cite{866330,1381917,1499041} assume the special case $K_B-1=K_{R}=M$. Using the shared buffer, we explore MARC transmission strategies where a UE can select either direct transmission or partial decode-and-forward transmission. 
Specifically, in a given slot one of the following three types of choices can be made: (A1) $\min \{ K_{B},M \}$ UEs transmit directly at rates that enable BS to decode, RS does not decode; (A2) $K_{R}$ UEs transmit at rates that enable RS to decode, while BS decodes them partially and waits for a subsequent cooperative message from the RS; (A3) RS forwards the side information to BS about the partially decoded messages, which are going to be combined and decoded entirely at the BS, while simultaneously $K_{B}-1$ UEs send new messages to the BS. For the proposed protocol, we formulate the associated optimization problems and derive the optimal transmission strategy. The numerical results show that the adaptive transmission selection can lead to substantial average throughput gains, when compared to fixed user scheduling combined with state-of-the-art approaches \cite{6284666,7039198}, as well as the outer bound of MARC with partial decode-and forward \cite{1381917}.

\section{System Model}
The system consists of $M$ UEs that act as sources $U_{m}, m=1,\ldots, M$, a RS as a relay $R$ with a shared buffer and a BS as a destination $B$, see Fig. \ref{model}. The buffer is assumed to be of an infinite size. Each UE has a sufficient backlog of messages to be sent. The transmission unit is a fixed length time slot. We assume that each node in the network sends with power $P$ and operates in half-duplex mode under a block fading channel. For each time slot, each $U_m$ has the channel state information (CSI) of the direct link $U_mB$ and the link to the relay $U_mR$. The CSI for all links $\{U_mB,U_mR\}$ is also available at the BS and the RS, along with the CSI of the link $RB$. In each slot one of the described three action types A1, A2, A3 take place. We assume that each data transmission slot is preceded by a negligibly short procedure for CSI acquisition. Since we assume block fading, the acquired CSI is valid throughout the data transmission slot. The decisions for the actions taken during the data transmission slot are made centrally at the RS and distributed to the UEs. Let $Q(i)$ denote the amount of normalized information in bits/symbol at the end of the $i$-th time slot in the relay's buffer.  Let $h_{XY}(i)$ denote the instantaneous channel coefficient of the link $XY$, $XY \in \{ U_mR,U_mB,RB,\}$, $\forall m \in \{1,\ldots,M \}$ at time slot $i$. The average channel gain of link $XY$ is given by $\Omega_{XY}=E \{  |h_{XY}(i)|^2 \}$, where $E\{\cdot\}$ denotes expectation. The instantaneous signal-to-noise ratio (SNR) with additive white Gaussian noise is given by $\gamma_{XY}(i) \stackrel{\triangle}{=} \frac{P|h_{XY}(i)|^2}{N_0}$. The instantaneous transmission rate of the UE $U_{m}$ and the relay $R$ are given by $R_{U_{m}}(i)$ and $R_{R}(i)$. We denote $ C(x) \stackrel{\triangle}{=} \log_{2} (1+x) $.

\section{Buffer-Aided Relaying Protocol for MARC}
\subsection{Instantaneous Transmission Schemes}
For fixed $K_{R}$ and $K_{B}$, there are $L=\binom{M}{ \min \{K_{B}, M\} }=\frac{M!}{\min \{K_{B}, M\}!(M-\min \{K_{B}, M\})!}$ possible subsets of active transmitters $\{ U_{k}^{(l)} | k \in \{ 1,\ldots,\min \{K_{B}, M\} \} \}$ in (A1). Let ${\cal S}_l$ be the $l-$th subset of $\min \{K_{B}, M\}$ UEs, where $l=1, 2, \ldots, L$. In (A2), the number of active transmitters is limited by $K_{R}$, such that there are $L^{\prime}=\binom{M}{ K_{R} }=\frac{M!}{K_{R}!(M-K_{R})!}$ possible subsets for picking $K_{R}$ UEs $\{ U_{k^{\prime}}^{(l^{\prime})} | k^{\prime} \in \{ 1,\ldots,K_{R} \} \}$. Let ${\cal S}_{l^{\prime}}^{\prime}$ be the $l^{\prime}-$th subset of $K_{R}$ UEs, where $l^{\prime}=1, 2, \ldots, L^{\prime}$. In (A3), since RS is one of the active transmitter, there are $L^{\prime \prime}=\binom{M}{K_{B}-1}=\frac{M!}{(K_{B}-1)!(M-K_{B}+1)!}$ possible subsets for picking the other $K_{B}-1$ transmitters $\{ U_{k^{\prime \prime}}^{(l^{\prime \prime})} | k^{\prime \prime} \in \{ 1,\ldots,K_{B}-1 \} \}$ from the UEs. Let ${\cal S}_{l^{\prime \prime}}^{\prime \prime}$ be the $l^{\prime \prime}-$th subset of $K_{B}-1$ UEs, where $l^{\prime \prime}=1, 2, \ldots, L^{\prime \prime}$. We use binary variables $p_{A1,l}(i)$, $p_{A2,l^{\prime}}(i), p_{A3,l^{\prime \prime}}(i) \in \{0,1\}$ to indicate whether in the $i-$th slot we have selected the transmission action A1, A2 and A3, respectively. Depending on the selected action, the rates are determined as follows:

\noindent \textbf{(A1):}  $p_{A1,l}(i)=1$, the rate region of the $\min \{ K_{B},M \}$ UEs from ${\cal S}_{l}$ should lie in the capacity region
\begin{IEEEeqnarray}{C}
\sum_{U_{k} \in {\cal W}} R_{U_{k}}(i) \leq C \left( \sum_{U_{k} \in {\cal W}} \gamma_{U_{k}B}(i) \right)  \notag
\end{IEEEeqnarray}
\noindent for every $ {\cal W} \subseteq {\cal S}_{l}$.
This is a polymatroid and the maximal sum-rate can be achieved by time-sharing, such that the sum-rate of the $\min \{ K_{B},M \}$ UEs are determined as 
\begin{IEEEeqnarray}{C}
\sum_{U_{k} \in {\cal S}_{l}} R_{U_{k}}(i) = C \left( \sum_{U_{k} \in {\cal S}_{l}} \gamma_{U_{k}B}(i) \right)  \notag
\end{IEEEeqnarray}
\noindent Since the RS is not used, the buffer state remains unchanged $Q(i)=Q(i-1)$.

\noindent \textbf{(A2):} $p_{A2,l^{\prime}}(i)=1$ happens only if $ \sum_{U_{k^{\prime}} \in {\cal W}^{\prime} } \gamma_{U_{k^{\prime}}R}(i) > \sum_{U_{k^{\prime}} \in {\cal W}^{\prime} } \gamma_{U_{k^{\prime}}B}(i) $ for all ${\cal W}^{\prime} \subseteq {\cal S}_{l^{\prime}}^{\prime}$. For further convenience, we denote indicator function 
\begin{IEEEeqnarray}{l}
{\cal X}_{A2,l^{\prime}}(i) \stackrel{\triangle}{=} {\cal X} \left( \sum_{U_{k^{\prime}} \in {\cal W}^{\prime} } \gamma_{U_{k^{\prime}}R}(i) > \sum_{U_{k^{\prime}} \in {\cal W}^{\prime} } \gamma_{U_{k^{\prime}}B}(i) \right) \notag \\
= \left\{\begin{array}{cl}
1, & \sum_{U_{k^{\prime}} \in {\cal W}^{\prime} } \gamma_{U_{k^{\prime}}R}(i) > \sum_{U_{k^{\prime}} \in {\cal W}^{\prime} } \gamma_{U_{k^{\prime}}B}(i) \notag \\
0, & \sum_{U_{k^{\prime}} \in {\cal W}^{\prime} } \gamma_{U_{k^{\prime}}R}(i) \leq \sum_{U_{k^{\prime}} \in {\cal W}^{\prime} } \gamma_{U_{k^{\prime}}B}(i) \notag
\end{array}
\right.
\end{IEEEeqnarray}
\noindent In words, the RS is only used if the sum-rate of the MAC at the RS is larger than the sum rate of the MAC at the BS. Here the sum-rate of the partially decoded messages at the BS is determined as:
\begin{IEEEeqnarray}{C}
\sum_{U_{k^{\prime}} \in {\cal S}_{l^{\prime}}^{\prime} } R_{U_{k^{\prime}}}^{ (1) }(i) = C \left( \sum_{U_{k^{\prime}} \in {\cal S}_{l^{\prime}}^{\prime} } \gamma_{U_{k^{\prime}}B}(i) \right)  \notag
\end{IEEEeqnarray}
Meanwhile, the remaining part of the messages is decoded at the RS with the sum-rate determined as
\begin{IEEEeqnarray}{C}
\sum_{U_{k^{\prime}} \in {\cal S}_{l^{\prime}}^{\prime} } R_{U_{k^{\prime}}}^{ (2)}(i) = C \left( \sum_{U_{k^{\prime}} \in {\cal S}_{l^{\prime}}^{\prime}} \gamma_{U_{k^{\prime}}R}(i) \right) - C \left( \sum_{U_{k^{\prime}} \in {\cal S}_{l^{\prime}}^{\prime} } \gamma_{U_{k^{\prime}}B}(i) \right) \notag
\end{IEEEeqnarray}
The RS stores cooperative messages, of rate $R_{U_{k^{\prime}}}^{ (2)}(i)$ 
\begin{IEEEeqnarray}{C}
Q(i)=Q(i-1)+ \sum_{U_{k^{\prime}} \in {\cal S}_{l^{\prime}}^{\prime} } R_{U_{k^{\prime}}}^{ (2)}(i) \notag
\end{IEEEeqnarray}

\noindent \textbf{(A3):} $p_{A3,l^{\prime \prime}}(i)=1$, $K_{B}-1$ UEs transmit new messages to the BS and the RS forwards the cooperative messages. The rate region of the $K_{B}-1$ UEs from ${\cal S}_{l^{\prime \prime}}^{\prime \prime}$ and RS lie in the capacity region:
\begin{IEEEeqnarray}{rCl}
\sum_{U_{k^{\prime \prime}} \in {\cal W}^{\prime \prime}} R_{U_{k^{\prime \prime}}}(i) & \leq & C \left( \sum_{U_{k^{\prime \prime}} \in {\cal W}^{\prime \prime}} \gamma_{U_{k^{\prime \prime}}B}(i) \right)  \notag \\
R_{R}(i) + \sum_{U_{k^{\prime \prime}} \in {\cal W}^{\prime \prime}} R_{U_{k^{\prime \prime}}}(i) & \leq & C \left( \gamma_{RB}(i) + \sum_{U_{k^{\prime \prime}} \in {\cal W}^{\prime \prime}} \gamma_{U_{k^{\prime \prime}}B}(i) \right)  \notag
\end{IEEEeqnarray}
\noindent for every $ {\cal W^{\prime \prime}} \subseteq {\cal S}_{l^{\prime \prime}}^{\prime \prime}$. If $K_{B}>1$, we impose that the RS uses a transmission rate such that its signal is the first one decoded by the BS, treating the other signals as interference. This is because the rate on RS-to-BS of (A3) has to be determined individually, since (A3) affects both directly the throughput and buffer state. It follows the intuition that once one of the possible (A3) is selected, the UEs should use the direct link at as high rate as possible to send new messages, i.e. the sum-rate of the UEs sending to the BS should not be diminished by the presence of the RS. In this way, the sum-rate of the $K_{B}-1$ UEs is determined as 
\begin{IEEEeqnarray}{C}
\sum_{U_{k^{\prime \prime}} \in {\cal S}_{l^{\prime \prime}}^{\prime \prime}} R_{U_{k^{\prime \prime}}}(i)  =  C \left( \sum_{U_{k^{\prime \prime}} \in {\cal S}_{l^{\prime \prime}}^{\prime \prime}} \gamma_{U_{k^{\prime \prime}}B}(i) \right)  \notag
\end{IEEEeqnarray}
With the additional constraint of the buffer state, the transmission rate over link $RB$ is 
\begin{IEEEeqnarray}{C}
R_{R}(i)= \min \left\{ Q(i-1), C \left(  \frac{ \gamma_{RB}(i) }{ 1+ \sum_{U_{k^{\prime \prime}} \in {\cal S}_{l^{\prime \prime}}^{\prime \prime}} \gamma_{U_{k^{\prime \prime}}B}(i) }  \right) \right\}  \notag
\end{IEEEeqnarray}
while the buffer is updated as $Q(i)= Q(i-1) - R_{R}(i)$.

\subsection{Optimal Transmission Strategy}
We derive the average throughput of the half-duplex MARC and provide the optimal transmission strategy to maximize the average system throughput. Since only one transmission mode is active in each slot, $\sum_{l=1}^{L} p_{A1,l}(i) + \sum_{l^{\prime}=1}^{L^{\prime}} p_{A2,l^{\prime}}(i)  + \sum_{l^{\prime \prime}=1}^{L^{\prime \prime}} p_{A3,l^{\prime \prime}}(i) =1$ has to be satisfied. The average arrival and departure rate in bits per slot to the buffer queue are:
\begin{IEEEeqnarray}{rl}
\bar{R}_{A} \stackrel{\triangle}{=} & \lim_{N \to \infty} \frac{1}{N} \sum_{i=1}^{N} \sum_{l^{\prime}=1}^{L^{\prime}}  p_{A2,l^{\prime}}(i) {\cal X}_{A2,l^{\prime}}(i) \sum_{U_{k^{\prime}} \in {\cal S}_{l^{\prime}}^{\prime} } R_{U_{k^{\prime}}}^{ (2) }(i) \\
\bar{R}_{D} \stackrel{\triangle}{=} & \lim_{N \to \infty} \frac{1}{N} \sum_{i=1}^{N} \sum_{l^{\prime \prime}=1}^{L^{\prime \prime}} p_{A3,l^{\prime \prime}}(i) R_{R}(i)
\end{IEEEeqnarray}

As the goal is to maximize the average throughput, the buffer should operate at the boundary of non-absorption, which can be proved rigorously, see \cite{6330084}. The buffer should be stable and in equilibrium we get:
\begin{IEEEeqnarray}{C}
\sum_{l^{\prime}=1}^{L^{\prime}} E \left\{ p_{A2,l^{\prime}}(i) {\cal X}_{A2,l^{\prime}}(i) \sum_{U_{k^{\prime}} \in {\cal S}_{l^{\prime}}^{\prime} } R_{U_{k^{\prime}}}^{ (2) }(i) \right\} = \sum_{l^{\prime \prime}=1}^{L^{\prime \prime}} E \left\{ p_{A3,l^{\prime \prime}}(i) R_{R}(i) \right\}
\label{protocol1C1}
\end{IEEEeqnarray}
with the corresponding average throughput (sum-rate):
\begin{IEEEeqnarray}{rl}
\bar{\tau} & = \frac{1}{N} \sum_{i=1}^{N} \Bigg\{ \sum_{l=1}^{L} p_{A1,l}(i) \sum_{U_{k} \in {\cal S}_{l}} R_{U_{k}}(i)  + \sum_{l^{\prime}=1}^{L^{\prime}} p_{A2,l^{\prime}}(i) {\cal X}_{A2,l^{\prime}}(i) \sum_{U_{k^{\prime}} \in {\cal S}_{l^{\prime}}^{\prime} } R_{U_{k^{\prime}}}^{ (1) }(i) \notag \\
& + \sum_{l^{\prime \prime}=1}^{L^{\prime \prime}} p_{A3,l^{\prime \prime}}(i) \Big[ R_{R}(i) + \sum_{U_{k^{\prime \prime}} \in {\cal S}_{l^{\prime \prime}}^{\prime \prime}} R_{U_{k^{\prime \prime}}}(i) \Big] \Bigg\}
\end{IEEEeqnarray}

The optimization problem for average throughput maximization can be formulated as
\begin{IEEEeqnarray}{rCl}
\max & & \bar{\tau} \label{opt1} \\
s.t. C1 &: &  \bar{R}_{A} = \bar{R}_{D} \notag \\
	 C2 &: & \sum_{l=1}^{L} p_{A1,l}(i) + \sum_{l^{\prime}=1}^{L^{\prime}} p_{A2,l^{\prime}}(i) + \sum_{l^{\prime \prime}=1}^{L^{\prime \prime}} p_{A3,l^{\prime \prime}}(i)  =1, \forall i \notag\\
	 C3 &: & p_{A1,l}(i), p_{A2,l^{\prime}}(i), p_{A3,l^{\prime \prime}}(i) \in \{ 0,1 \},\forall l,l^{\prime},l^{\prime \prime},i \notag
\end{IEEEeqnarray}
where C1 ensures that (\ref{protocol1C1}) holds, C2 and C3 ensure that a single action (A1), (A2) or (A3) is selected in each slot.

In the optimization problem \eqref{opt1}, the variables we need to optimize include binary indicators for the candidate actions in each slot with coupled queue state. Hence, the first step is to decouple the buffer state from CSI, thus $\min \{\cdot\}$ function is eliminated, since the event that (A3) is selected while the output of the buffer is limited by $Q(i-1)$ is negligible over a long time $N \to \infty$, as also done in \cite{6330084,6666121,6284666}. This implies that we are dealing with a $0-1$ integer linear programming problem. To offer a tractable solution, we relax the binary constraints to the closed interval $[0,1]$ which reveals that the feasible set of the problem is enlarged. However, the possible solution of the relaxed problem lies on the boundary, and in fact it is solution of the original problem. The relaxed problem is solved by Lagrange multipliers and the KKT conditions.

\begin{algorithm}[!t]
\caption{Gradient Algorithm for $\lambda^{*}$ }\label{ga1}
\begin{algorithmic}[1]
\State \textbf{initialize} $s=0$, $\lambda[0]$
\Repeat
	\State Compute $p_{A1,l}^{*}(i), p_{A2,l^{\prime}}^{*}(i), p_{A3,l^{\prime \prime}}^{*}(i), \forall i$ according to Proposition 1
	\State Compute $\Delta \lambda[s]$ based on \eqref{delta1}
	\State Update $\lambda[s+1]$ based on \eqref{update1}
	\State $s \gets s+1$
\Until{converge to $\lambda^{*}$}
\end{algorithmic}
\end{algorithm}

\begin{myprop}
The optimal decision functions for maximizing the average throughput with buffer-aided relaying protocol are:

Case I: If $\lambda > -1$, the criterion is 
\begin{IEEEeqnarray}{lCl}
p_{A1,l}^{*}(i) &=&
\left\{
\begin{array}{lcl}
1, & \mbox{if}    & \Lambda_{A1,l}(i) \geq \Lambda_{A1,j}(i), \forall j \neq l \\
    & \mbox{and} & \Lambda_{A1,l}(i) \geq \Lambda_{A2,j}(i), \forall j \\
    & \mbox{and} & \Lambda_{A1,l}(i) \geq \Lambda_{A3,j}(i), \forall j \\
0, &            & \mbox{otherwise}
\end{array}
\right.
\label{proposition1a}\\
p_{A2,l^{\prime}}^{*}(i) &=&
\left\{
\begin{array}{lcl}
1, & \mbox{if}    & \Lambda_{A2,l^{\prime}}(i) \geq \Lambda_{A2,j}(i), \forall j \neq l^{\prime} \\
    & \mbox{and} & \Lambda_{A2,l^{\prime}}(i) \geq \Lambda_{A1,j}(i), \forall j \\
    & \mbox{and} & \Lambda_{A2,l^{\prime}}(i) \geq \Lambda_{A3,j}(i), \forall j  \\
0, &            & \mbox{otherwise}
\end{array}
\right.
\label{proposition1b}\\
p_{A3,l^{\prime \prime}}^{*}(i) &=&
\left\{
\begin{array}{lcl}
1, & \mbox{if}    & \Lambda_{A3,l^{\prime \prime}}(i) \geq \Lambda_{A3,j}(i), \forall j \neq l^{\prime \prime} \\
    & \mbox{and} & \Lambda_{A3,l^{\prime \prime}}(i) \geq \Lambda_{A1,j}(i), \forall j \\
    & \mbox{and} & \Lambda_{A3,l^{\prime \prime}}(i) \geq \Lambda_{A2,j}(i), \forall j \\
0, &            & \mbox{otherwise}
\end{array}
\right.
\label{proposition1c}
\end{IEEEeqnarray}
Case II: If $\lambda \leq -1$, the criterion is 
\begin{IEEEeqnarray}{rCl}
p_{A1,l}^{*}(i) &=&
\left\{
\begin{array}{lcl}
1, & \mbox{if}    & \Lambda_{A1,l}(i) \geq \Lambda_{A1,j}(i), \forall j \neq l  \\
0, &            & \mbox{otherwise}
\end{array}
\right.
\end{IEEEeqnarray}
where selection matrices are denoted by
\begin{IEEEeqnarray}{lCl}
\Lambda_{A1,l} (i) &=& C \left( \sum_{U_{k} \in {\cal S}_{l}} \gamma_{U_{k}B}(i) \right), \forall l \notag \\
\Lambda_{A2,l^{\prime}} (i) &=& {\cal X}_{A2,l^{\prime}}(i) \Big[ (1+\lambda) C \Big( \sum_{U_{k^{\prime}} \in {\cal S}_{l^{\prime}}^{\prime} } \gamma_{U_{k^{\prime}}B}(i) \Big) - \lambda C \Big( \sum_{U_{k^{\prime}} \in {\cal S}_{l^{\prime}}^{\prime} } \gamma_{U_{k^{\prime}}R}(i) \Big) \Big], \forall l^{\prime}  \notag \\
\Lambda_{A3,l^{\prime \prime}} (i) &=& C \Big( \sum_{U_{k^{\prime \prime}} \in {\cal S}_{l^{\prime \prime}}^{\prime \prime} } \gamma_{U_{k^{\prime \prime}}B}(i) \Big) + (1+\lambda) C \Big( \frac{ \gamma_{RB}(i) }{ 1+ \sum_{U_{k^{\prime \prime}} \in {\cal S}_{l^{\prime \prime}}^{\prime \prime}} \gamma_{U_{k^{\prime \prime}}B}(i) } \Big), \forall l^{\prime \prime}  \notag
\end{IEEEeqnarray}
and $\lambda$ denotes the Lagrange multiplier corresponding to C1. 
\end{myprop}

In Case I $\lambda > -1$, the optimal $\lambda$ under fading can be obtained numerically and iteratively with gradient algorithm using the following update equation.
\begin{IEEEeqnarray}{C}
\lambda[s+1]=\lambda[s] + \delta[s] \Delta \lambda[s] \label{update1}
\end{IEEEeqnarray}
where $s$ is the iteration index and $\delta [s]$ is step size which has to be chosen appropriately. In each iteration, the optimal decision indicators are obtained according to Proposition 1 and then the following expression updates as
\begin{IEEEeqnarray}{rl}
\Delta \lambda[s] &= \sum_{l^{\prime}=1}^{L^{\prime}} E \Big\{ p_{A2,l^{\prime}}^{*}(i)  {\cal X}_{A2,l^{\prime}}(i)  \Big[ C \Big( \sum_{U_{k^{\prime}} \in {\cal S}_{l^{\prime}}^{\prime} } \gamma_{ U_{k^{\prime}}R }(i)  \Big) \notag \\
& - C \Big( \sum_{U_{k} \in {\cal S}_{l^{\prime}}^{\prime} } \gamma_{U_{k^{\prime}}B}(i)  \Big)  \Big] \Big\}  \notag \\
& - \sum_{l^{\prime \prime}=1}^{L^{\prime \prime}} E \Big\{ p_{A3,l^{\prime \prime}}^{*}(i) C \Big( \frac{ \gamma_{RB}(i) }{ 1+ \sum_{U_{k^{\prime \prime}} \in {\cal S}_{l^{\prime \prime}}^{\prime \prime}} \gamma_{U_{k^{\prime \prime}}B}(i) } \Big) \Big\} 
\label{delta1}
\end{IEEEeqnarray}
We summarize the numerical approach in Algorithm 1.

In Case II $\lambda \leq -1$, there is no need for RS to aid the communication, but only selection of the transmission mode that has the maximal sum-rate to the BS from (A1), considering the access limit by BS.

\section{Simulation Results}
We present simulation results to compare the performance of the proposed method with state-of-the-art protocols for buffer-aided relaying from \cite{6284666} and \cite{7039198}, which are applied to a multi-user system by using round-robin scheduling. We also compare to the outer bound of MARC from \cite{1381917}, which requires $K_{R}=M$ and $K_{B}=M+1$.

We set $M=3$ UEs and denote $\Gamma=\frac{P}{N_{0}}$. All links are subject to Rayleigh fading. All schemes assume the use of an infinite buffer. We denote the average channel gain vector of all the involved links as $\mathbf{\Omega}=[ \Omega_{U_{1}R}, \Omega_{U_{2}R}, \Omega_{U_{3}R}, \Omega_{U_{1}B}, \Omega_{U_{2}B}, \Omega_{U_{3}B}, \Omega_{RB}]$.

\begin{figure}[!t]
\centering
\includegraphics[width=3in]{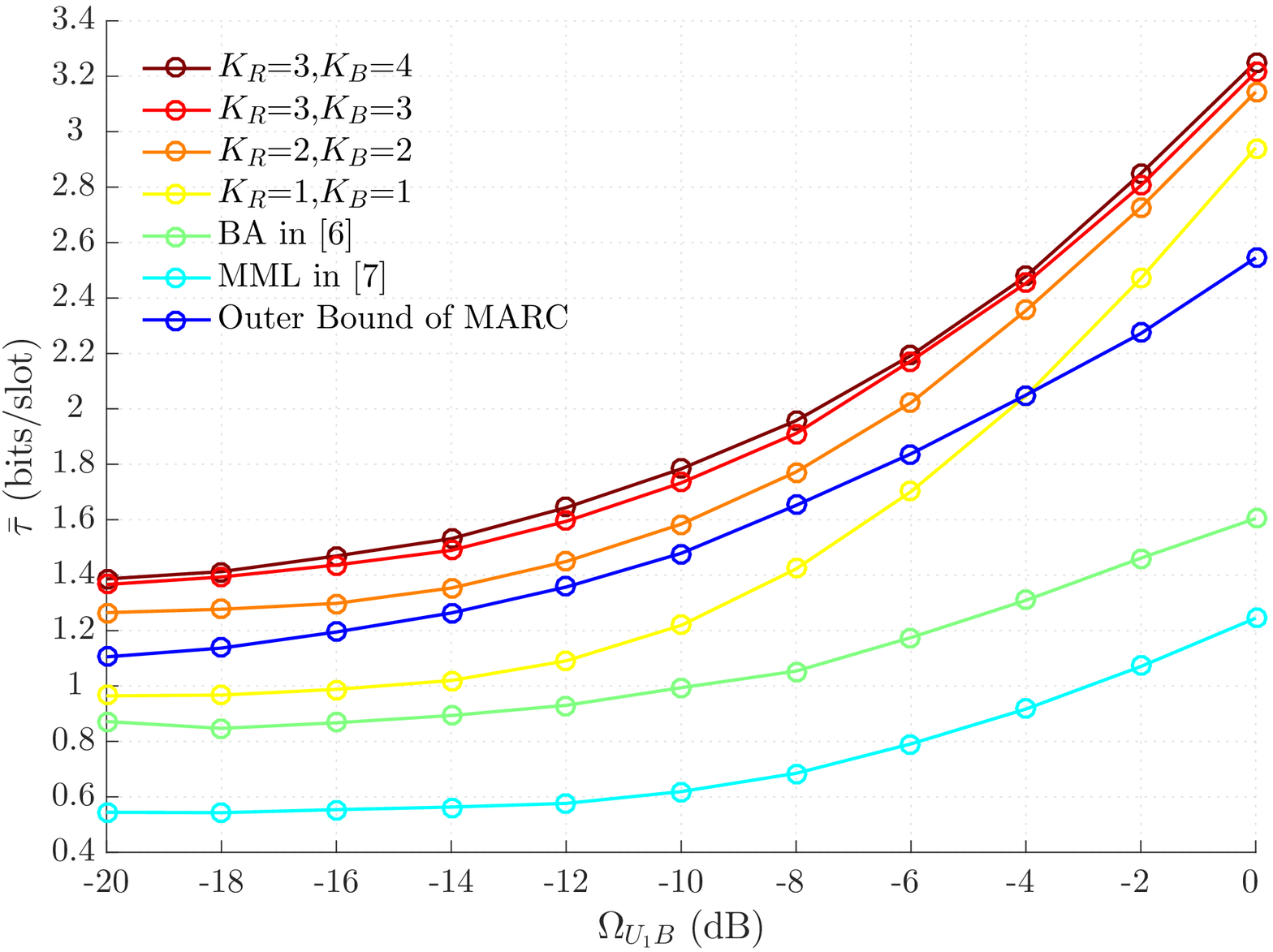}
\caption{Average throughput vs. $\Omega_{U_{1}B}$ for $\Gamma=10$dB,  $\mathbf{\Omega} = [ -11, -9, -8,$ $\Omega_{U_{1}B},-13, -15, -10 ]$dB.}
\label{simfig2}
\end{figure}

In Fig. \ref{simfig2}, we show the effect of the average throughput with respect to $\Omega_{U_{1}B}$. Our proposed buffer-aided relaying protocol shows a better performance than BA in \cite{6284666} and MML in \cite{7039198}. Moreover, increasing $K_{R}$ or $K_{B}$ can enhance the average system throughput. There is a slow growth when $\Omega_{U_{1}B} < -12$dB and a fast growth when $\Omega_{U_{1}B} > -12$dB. This is because as the average channel gain $\Omega_{U_{1}B}$ becomes better, the probability of the direct selection transmission modes involving link $U_{1}B$ goes higher. Further in the general case $K_{R}>1$ and $K_{B}>1$, our proposed protocol outperforms the outer bound of the MARC. There is even a cross point between the case $K_{R}=K_{B}=1$ and outer bound of MARC, which means that even the RS and BS could not deal with the multiuser decoding, the proposed protocol may benefit from the direct transmission. 

\begin{figure}[!t]
\centering
\includegraphics[width=3in]{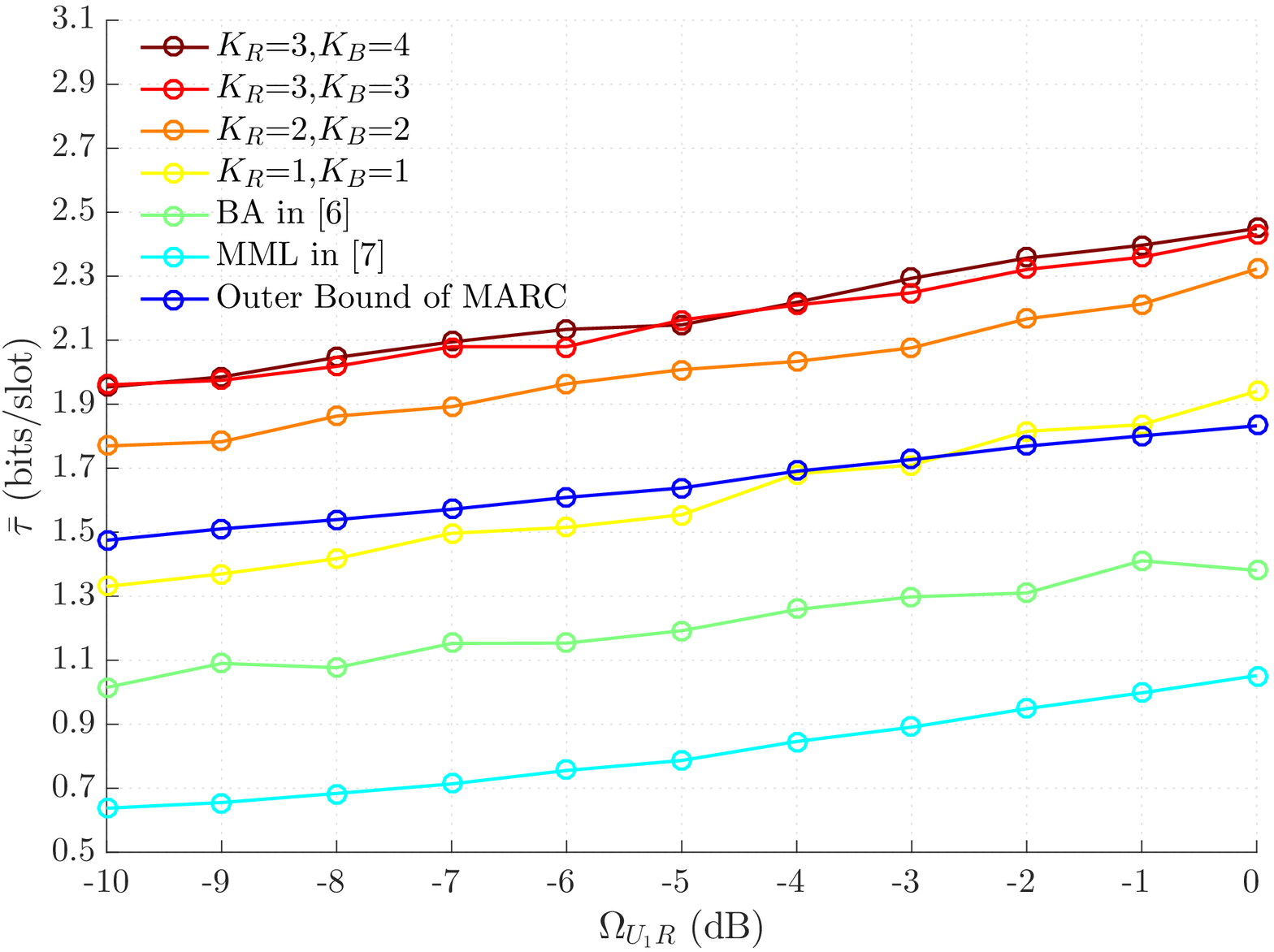}
\caption{Average throughput vs. $\Omega_{U_{1}R}$ for $\Gamma=10$dB, $\mathbf{\Omega}=[\Omega_{U_{1}R}, -9, -8,$ $-16, -13, -15, 0]$dB.}
\label{simfig4}
\end{figure}

Fig. \ref{simfig4} shows the performance with respect to $\Omega_{U_{1}R}$. In this scenario, each direct link is on average weaker than the UE-to-RS links and the backhaul link (RS-to-BS) is strong in order to guarantee a high rate of the relayed transmission. The proposed protocols achieve large gains and the average throughput increases with $\Omega_{U_{1}R}$.

\section{Conclusion}
We have introduced a class of transmission protocols that can adaptively select between direct and relayed uplink transmission in a scenario with multiple UEs. The proposed protocols outperform the state-of-the-art approaches that use buffer-aided relaying. They are also better than the outer bound of MARC with partial decode-and-forward strategy when  $K_{R}>1$ and $K_{B}>1$, since we are simultaneously reaping the benefits of multiuser diversity and adaptive transmit mode. An item for further work is the impact of imperfect CSI, where the imperfectness varies with the link type.

\bibliographystyle{IEEEtran}
\bibliography{mybibfile}
\end{document}